\pdfoutput=1 
\documentclass[cernpreprint,texlive=2011,UKenglish,texmf]{atlasdoc}

\usepackage[biblatex=true,backend=biber]{atlaspackage}

\usepackage[articletitle=false]{atlasbiblatex}

\usepackage{atlasphysics}

\usepackage{BcDsJpsiPaper-defs}

\AtlasTitle{Study of the $B_c^+ \to J/\psi D_s^+$
and $B_c^+ \to J/\psi D_s^{*+}$ decays with the ATLAS 
detector}

\author{The ATLAS Collaboration}

\hypersetup{pdftitle={Study of the \texorpdfstring{$B_c^+ \to J/\psi 
D_s^+$}{Bc+ -> J/psi Ds+}
and \texorpdfstring{$B_c^+ \to J/\psi D_s^{*+}$}{Bc+ -> J/psi Ds*+} decays with 
the ATLAS 
detector},pdfauthor={The ATLAS Collaboration}}

\PreprintIdNumber{CERN-PH-2015-163}

\AtlasJournal{Eur.\ Phys.\ J.\ C}

\AtlasAbstract{%
The decays $B_c^+ \to J/\psi D_s^+$ and $B_c^+ \to J/\psi D_s^{*+}$ are studied 
with the ATLAS detector at the LHC using a dataset corresponding to integrated 
luminosities of 4.9 fb$^{-1}$ and 20.6 fb$^{-1}$ of $pp$ collisions collected at 
centre-of-mass energies $\sqrt{s} = 7$ TeV and $8$ TeV, respectively. Signal 
candidates are identified through $J/\psi\to\mu^+\mu^-$ and 
$D_s^{(*)+}\to\phi\pi^+(\gamma/\pi^0)$ decays. With a two-dimensional likelihood 
fit involving the $B_c^+$ reconstructed invariant mass and an angle between the 
$\mu^+$ and $D_s^+$ candidate momenta in the muon pair rest frame, the yields of 
$B_c^+ \to J/\psi D_s^+$ and $B_c^+ \to J/\psi D_s^{*+}$, and the transverse 
polarisation fraction in $B_c^+ \to J/\psi D_s^{*+}$ decay are measured. The 
transverse polarisation fraction is determined to be $\Gamma_{\pm\pm}(B_c^+ \to 
J/\psi D_s^{*+})/\Gamma(B_c^+ \to J/\psi D_s^{*+}) = 0.38 \pm 0.23 \pm 0.07$, 
and the derived ratio of the branching fractions of the two modes is 
$\mathcal{B}_{B_c^+ \to J/\psi D_s^{*+}}/\mathcal{B}_{B_c^+ \to J/\psi D_s^+} = 
2.8 \,^{+1.2}_{-0.8} \pm 0.3$, where the first error is statistical and the 
second is systematic. Finally, a sample of $B_c^+\to J/\psi\pi^+$ decays is used 
to derive the ratios of branching fractions $\mathcal{B}_{B_c^+ \to J/\psi 
D_s^+}/\mathcal{B}_{B_c^+ \to J/\psi \pi^+} = 3.8 \pm 1.1 \pm 0.4 \pm 0.2$ and 
$\mathcal{B}_{B_c^+ \to J/\psi D_s^{*+}}/\mathcal{B}_{B_c^+ \to J/\psi \pi^+} = 
10.4 \pm 3.1 \pm 1.5 \pm 0.6$, where the third error corresponds to the 
uncertainty of the branching fraction of $D_s^+\to\phi(K^+K^-)\pi^+$ decay. The 
available theoretical predictions are generally consistent with the measurement.
}

\begin{document}

\maketitle

\section{Introduction}\label{sec:Intro}
\begin{figure*}[htbp]
\centering
\subfloat[]{
\includegraphics[width=0.3\textwidth]{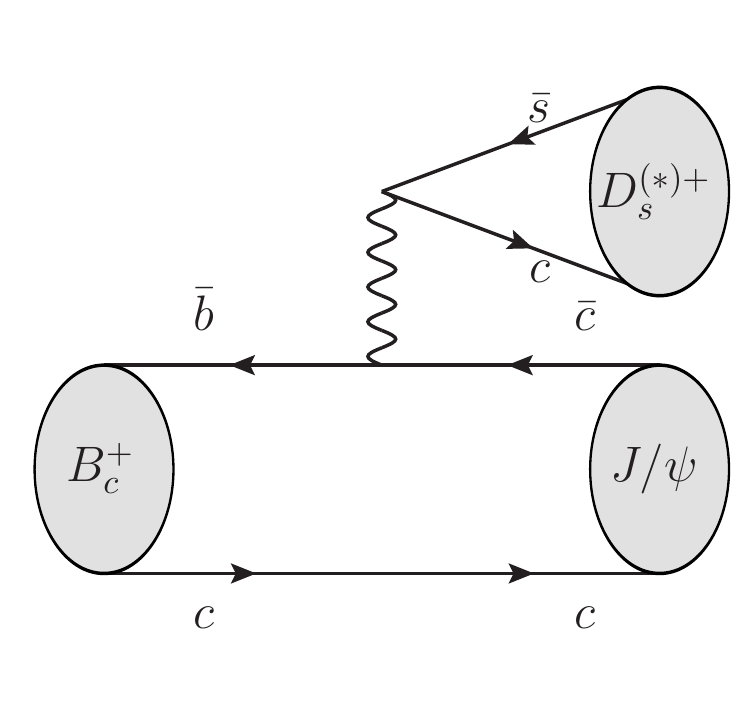}
\label{fig:Diagrams:Diag_spectator}
}
\subfloat[]{
\includegraphics[width=0.3\textwidth]{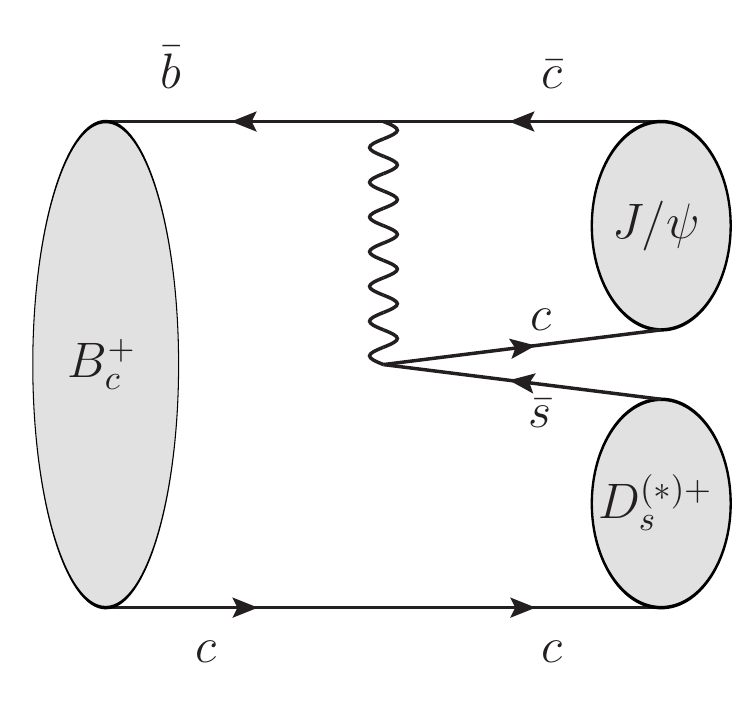}
\label{fig:Diagrams:Diag_spectator_colsup}
}
\subfloat[]{
\includegraphics[width=0.3\textwidth]{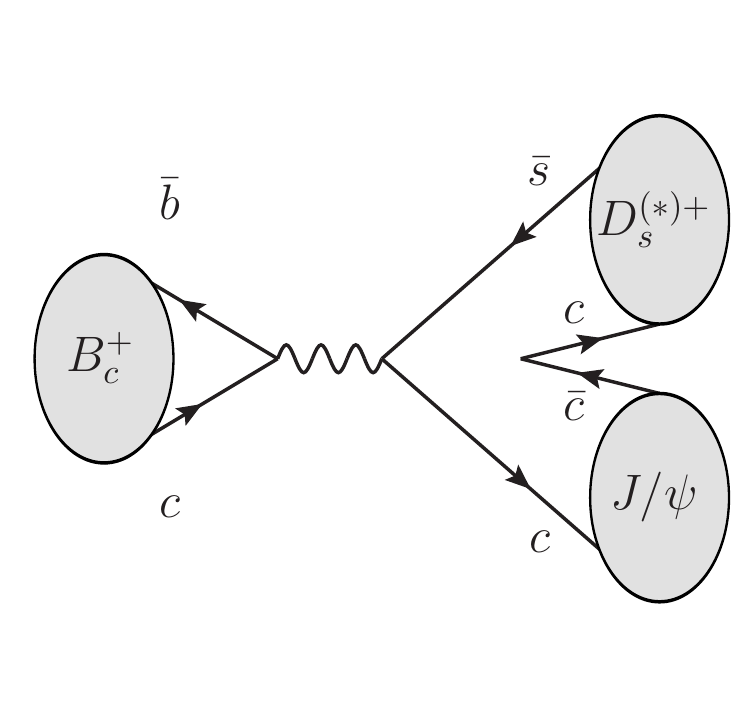}
\label{fig:Diagrams:Diag_annihil}
}
\caption{Feynman diagrams for $\Bc\to\Jpsi\DsOrStar$ decays: 
\protect\subref{fig:Diagrams:Diag_spectator} colour-favoured spectator, 
\protect\subref{fig:Diagrams:Diag_spectator_colsup} colour-suppressed 
spectator, and
\protect\subref{fig:Diagrams:Diag_annihil} annihilation topology.
\label{fig:Diagrams}}
\end{figure*}

The $\Bc$ meson%
\footnote{Charge conjugate states 
are implied throughout the paper unless otherwise stated.} 
is the only known weakly decaying particle consisting of two 
heavy quarks. The ground $\bar{b}c$ state was first observed by 
CDF~\cite{Abe:1998wi} via its
semileptonic decay $\Bc\to\Jpsi\ell^+\nu_\ell$. 
An excited $\bar{b}c$ state has been observed recently by 
ATLAS~\cite{Aad:2014laa} using the $\Bc$ decay mode 
$\Bc\to\Jpsi\pi^+$. The presence of two heavy quarks, 
each of which can decay weakly, affects theoretical calculations of the decay 
properties of the $\Bc$ meson. 
In the case of $\bar{b}\to \bar{c}c\bar{s}$ processes, decays to charmonium 
and a $\Ds$ or a $\DsStar$ meson are predicted to occur via colour-suppressed
and colour-favoured spectator diagrams as well as via the weak annihilation 
diagram 
(see Fig.~\ref{fig:Diagrams}). The latter, in contrast to decays of other 
$B$ mesons, is not Cabibbo-suppressed and can contribute significantly to the 
decay amplitudes. The decay properties are addressed in various
theoretical calculations~\cite{Colangelo:1999zn, Kiselev:2002vz, Ivanov:2006ni, 
Dhir:2008hh, Ke:2013yka, Rui:2014tpa, Kar:2013fna} and can also be compared to 
the analogous properties in the lighter $B$ meson systems such as $\Bd\to 
D^{*-}\DsOrStar$ or $B^+\to\bar{D}^{*0}\DsOrStar$. 
The decays $\Bc\to\Jpsi\Ds$ and $\Bc\to\Jpsi\DsStar$, which have been 
observed recently by the LHCb experiment~\cite{Aaij:2013gia}, provide a means to 
test these theoretical predictions.

This paper 
presents a measurement of the branching fractions of $\Bc\to\Jpsi\Ds$ 
and $\Bc\to\Jpsi\DsStar$ decays, normalised to that of $\Bc\to\Jpsi\pi^+$ 
decay, 
and polarisation in $\Bc\to\Jpsi\DsStar$ decay performed
with the ATLAS detector~\cite{Aad:2008zzm}. 
The $\Ds$ meson is reconstructed
via the $\Ds\to\phi\pi^+$ decay with the $\phi$ meson decaying into a pair of 
charged 
kaons. 
The $\DsStar$ meson decays into a $\Ds$ meson and a soft photon or $\pi^0$. 
Detecting such soft neutral particles is very challenging, 
thus no attempt to reconstruct them is made in the analysis.
The $\Jpsi$ meson is reconstructed via its decay into a muon pair. 

The measurement presented in this paper allows an independent verification of the results of Ref.~\cite{Aaij:2013gia} with comparable statistical and systematic uncertainties.
The following ratios are measured: $\R_{\Ds/\pi^+} = 
\Br_{\Bc\to\Jpsi\Ds}/\Br_{\Bc\to\Jpsi\pi^+}$, $\R_{\DsStar/\pi^+} = 
\Br_{\Bc\to\Jpsi\DsStar}/\Br_{\Bc\to\Jpsi\pi^+}$, and $\R_{\DsStar/\Ds} = 
\Br_{\Bc\to\Jpsi\DsStar}/\Br_{\Bc\to\Jpsi\Ds}$, where $\Br_{\Bc\to X}$ denotes 
the branching fraction of the $\Bc\to X$ decay.
The decay $\Bc\to\Jpsi\DsStar$ is a transition of a pseudoscalar meson into 
a pair of vector states and is thus described by the three helicity 
amplitudes, $A_{++}$, $A_{--}$, and $A_{00}$, where the subscripts correspond 
to the helicities of $\Jpsi$ and $\DsStar$ mesons. The contribution of the 
$A_{++}$ and $A_{--}$ amplitudes, referred to as the $A_{\pm\pm}$ component,
corresponds to the $\Jpsi$ and $\DsStar$ transverse polarisation. The fraction 
of transverse polarisation, 
$\Gamma_{\pm\pm}/\Gamma = \Gamma_{\pm\pm}
(\Bc\to\Jpsi\DsStar)/\Gamma(\Bc\to\Jpsi\DsStar)$, is also 
measured. From a naive prediction by spin counting, one would expect this 
fraction to be $2/3$, while calculations~\cite{Rui:2014tpa,Kar:2013fna} predict values of 0.41--0.48.

This analysis is based on a combined sample of $pp$ collision 
data collected by the ATLAS experiment at the LHC at centre-of-mass energies 
$\sqrt{s} = 7$\,\TeV\ and 
$8$\,\TeV\ corresponding to integrated luminosities of 4.9~\ifb\ and 
20.6~\ifb, respectively.

\section{The ATLAS detector, trigger selection and Monte Carlo 
samples}\label{sec:ATLAS}
ATLAS is a general-purpose detector  
consisting of several subsystems including the inner detector (ID), 
calorimeters and the muon spectrometer (MS). 
Muon reconstruction makes use of both the ID and the MS.
The ID comprises three types of detectors: a 
silicon pixel detector, a silicon microstrip semiconductor tracker (SCT) and a 
transition radiation tracker. The ID provides a 
pseudorapidity%
\footnote{ATLAS uses a right-handed coordinate system with its origin at the 
nominal interaction point (IP) in the centre of the detector and the $z$-axis 
along the beam pipe. The $x$-axis points from the IP to the centre of the LHC 
ring, and the $y$-axis points upward. Cylindrical coordinates $(r,\phi)$ are 
used in the transverse plane, $\phi$ being the azimuthal angle around the beam 
pipe. The pseudorapidity is defined in terms of the polar angle $\theta$ as 
$\eta=-\ln\tan(\theta/2)$.} 
coverage up to 
$|\eta| = 2.5$. Muons pass through the calorimeters and reach the MS if their 
transverse momentum, $\pT$, is above approximately 3\,\GeV.%
\footnote{Using a system of units with $c = 1$ is implied throughout the paper.}
Muon candidates are 
formed 
either from a stand-alone MS track matched to an ID track or, in case the MS 
stand-alone track is not reconstructed, from 
an ID track extrapolated to the MS and matched to 
track segments in the MS.
Candidates of the latter type are referred to as segment-tagged muons while the 
former are called combined muons.
Muon track parameters are taken from 
the ID measurement
alone in this analysis, since the precision of the measured track parameters 
for muons in the $\pT$ range of interest is dominated by the ID track 
reconstruction.

The ATLAS trigger system consists of a 
hardware-based Level-1 trigger and a two-stage High Level Trigger (HLT). At 
Level-1, the muon 
trigger uses dedicated MS chambers to search for patterns of hits satisfying 
different $\pT$ thresholds. The region-of-interest around these 
hit patterns then serves as a seed for the HLT muon reconstruction, in which 
dedicated algorithms are used to incorporate information from both the MS and 
the ID, achieving a position and momentum resolution close to that provided by 
the offline muon reconstruction. Muons are efficiently triggered in 
the pseudorapidity range $|\eta| < 2.4$.

Triggers based on single-muon, dimuon, and three-muon signatures are 
used to select $\Jpsi\to\mu^+\mu^-$ decays for the analysis. The third muon can 
be produced in the $\Bc$ signal events in semileptonic decays of the two other 
heavy-flavour hadrons. The majority of events are collected by dimuon 
triggers requiring a vertex of two oppositely charged muons with an invariant 
mass 
between 2.5\,\GeV\ and 4.3\,\GeV. During the data taking, the $\pT$ threshold 
for muons in these triggers was either 4\,\GeV\ or 6\,\GeV.
Single-muon triggers additionally increase the acceptance for 
asymmetric $\Jpsi$ 
decays where one muon has $\pT 
< 4$\,\GeV. Finally, three-muon triggers had a $\pT$ threshold of 4\,\GeV, 
thus enhancing the acceptance
during the periods of high luminosity when the $\pT$ threshold for at least one 
muon in the dimuon 
triggers was 6\,\GeV. 

Monte Carlo (MC) simulation is used for the event selection criteria 
optimisation and the calculation of the acceptance for the considered $\Bc$ 
decay modes. The MC samples of the $\Bc$ decays were generated with 
\textsc{Pythia}~6.4~\cite{Sjostrand:2006za} along with a dedicated extension 
for the $\Bc$ production based on calculations from 
Refs.~\cite{Berezhnoy:1995au, Berezhnoy:1996ks, 
Berezhnoy:1997fp, Berezhnoy:2004gc}. 
The decays of $\Bc$ are 
then simulated with \textsc{EvtGen}~\cite{Lange:2001uf}.
The generated events were  
passed through a full simulation of the detector using the ATLAS 
simulation framework~\cite{Aad:2010ah} based on 
\textsc{Geant~4}~\cite{Agostinelli:2002hh, 
Allison:2006ve} and processed with the same reconstruction algorithms as 
were used for the data.

\section{Reconstruction and event selection}\label{sec:Reco}
The $\Jpsi$ candidates are reconstructed from pairs of oppositely charged 
muons. At 
least one of the two muons is required to be a combined muon. Each pair is 
fitted to a 
common vertex~\cite{Kostyukhin:2003daa}. The quality of the vertex fit must 
satisfy $\chi^2/\ndf < 15$, where the $\ndf$ stands for the number of degrees 
of freedom.
The candidates in the invariant mass window $2800\MeV < m(\mu^+\mu^-) < 
3400$\,\MeV\ are retained. 

For the $\Ds\to\phi(K^+K^-)\pi^+$ reconstruction, tracks of particles with 
opposite charges are assigned kaon mass hypotheses and combined in pairs to 
form $\phi$ candidates. An 
additional track is assigned a pion mass and combined with the $\phi$ 
candidate to form a $\Ds$ candidate. 
To ensure good momentum resolution, all three tracks are required to have at 
least two hits in the silicon pixel detector
and at least six hits in the SCT.
Only three-track combinations successfully fitted to 
a common vertex with $\chi^2/\ndf < 8$ are kept. 
The $\phi$ candidate invariant mass, $m(K^+K^-)$, and the $\Ds$  candidate 
invariant mass, $m(K^+K^-\pi^+)$, are calculated 
using the track momenta refitted to the common vertex. Only candidates with 
$m(K^+K^-)$ within $\pm 7$\,\MeV\ around the $\phi$ 
mass, $m_{\phi} = 1019.461$\,\MeV~\cite{Agashe:2014kda}, and with $1930\MeV < 
m(K^+K^-\pi^+) < 2010$\,\MeV\ are retained.

The $\Bc\to\Jpsi\Ds$ candidates are built by 
combining the five tracks of the $\Jpsi$ and $\Ds$ candidates.
The $\Jpsi$ meson decays instantly at the same point as the $\Bc$ does 
(secondary vertex) 
while the $\Ds$ lives long enough to form a displaced tertiary vertex. Therefore 
the five-track combinations are refitted assuming this cascade topology~\cite{Kostyukhin:2003daa}. The 
invariant mass of the muon pair is constrained to the $\Jpsi$\ 
mass, $m_{\Jpsi} = 3096.916$\,\MeV~\cite{Agashe:2014kda}. The three $\Ds$ 
daughter tracks are constrained to a tertiary vertex and their invariant
mass is fixed to the mass of $\Ds$, $m_{\Ds} = 
1968.30$\,\MeV~\cite{Agashe:2014kda}. The 
combined momentum of the refitted 
$\Ds$ decay tracks is constrained to point to the dimuon vertex. 
The quality of the cascade fit must satisfy $\chi^2/\ndf < 3$. 

The $\Bc$ meson is reconstructed within the kinematic range $\pT(\Bc) > 
15$\,\GeV\ and $|\eta(\Bc)| < 
2.0$, where the detector acceptance is high and depends weakly on $\pT(\Bc)$ 
and $\eta(\Bc)$.

The refitted tracks of the $\Ds$ daughter hadrons are required to have $|\eta| < 
2.5$ and $\pT > 1$\,\GeV, while the muons must have $|\eta| < 2.3$ and $\pT > 
3$\,\GeV. 
To further discriminate the sample of $\Ds$ 
candidates from a large combinatorial background, the following requirements 
are applied:
\begin{itemize}
\item $\cos\theta^*(\pi) < 0.8$, where $\theta^*(\pi)$ is the angle between the 
pion momentum in the $K^+K^-\pi^+$ rest frame and the $K^+K^-\pi^+$ combined 
momentum in the laboratory frame;
\item $|\cos^3\theta^\prime(K)| > 0.15$, where $\theta^\prime(K)$ is the angle 
between one of the kaons and the pion in the $K^+K^-$ rest frame. The decay of 
the 
pseudoscalar $\Ds$ meson to the $\phi$ (vector) plus $\pi$ (pseudoscalar) final 
state results in an alignment of the spin of the $\phi$ meson perpendicularly 
to the direction of motion of the $\phi$ relative to $\Ds$. Consequently, the 
distribution of $\cos\theta^\prime(K)$ follows a $\cos^2\theta^\prime(K)$ 
shape, implying a uniform distribution for $\cos^3\theta^\prime(K)$. In 
contrast, 
the $\cos\theta^\prime(K)$ distribution of the combinatorial background is 
uniform 
and its $\cos^3\theta^\prime(K)$ distribution peaks at zero. The cut suppresses 
the background significantly while reducing the signal by 15\%.
\end{itemize}

The $\Bc$ candidate is required to point back to a primary vertex such that 
$d_0^{\rm PV}(\Bc) < 0.1$~mm and $z_0^{\rm PV}(\Bc)\sin\theta(\Bc) < 0.5$~mm, 
where $d_0^{\rm PV}$ and $z_0^{\rm PV}$ are respectively the transverse and 
longitudinal impact parameters with respect to the primary vertex. All primary 
vertices in the event are considered. If there 
is more than one primary vertex satisfying these requirements ($\sim 0.5$\% 
events both in data and MC simulation), the one with the largest sum of squared 
transverse 
momenta of the tracks originating from it is chosen.

The transverse decay length%
\footnote{The transverse decay length of a particle is 
defined as the transverse distance between the production (primary) vertex 
and the particle decay (secondary) vertex projected along its transverse 
momentum.} 
of the $\Bc$ candidate is required to satisfy 
$L_{xy}(\Bc) > 0.1$~mm. The transverse decay length of the $\Ds$ 
measured from the $\Bc$ vertex must be $L_{xy}(\Ds) > 0.15$~mm. In order to 
remove fake candidates, both 
$L_{xy}(\Bc)$ and $L_{xy}(\Ds)$ are required not to exceed 10~mm.

Taking into account the characteristic hard fragmentation of $b$-quarks, a 
requirement 
$\pT(\Bc)/\sum\pT(\mathrm{trk}) > 0.1$ is applied, where the sum in the 
denominator 
is taken over all tracks originating from the primary vertex (tracks of the $\Bc$ candidate are included in the sum if they are associated with the primary vertex). The requirement 
reduces a sizeable fraction of combinatorial background while having almost no 
effect on the signal.

The following angular selection requirements are introduced to further suppress 
the combinatorial background:
\begin{itemize}
\item $\cos\theta^*(\Ds) > -0.8$, where $\theta^*(\Ds)$ is the angle between 
the $\Ds$ candidate momentum in the 
rest frame of the $\Bc$ candidate, and the $\Bc$ candidate line of flight in 
the laboratory frame. The distribution of $\cos\theta^*(\Ds)$ is uniform for the
decays of pseudoscalar $\Bc$ meson before any kinematic selection while it tends 
to increase for negative values of $\cos\theta^*(\Ds)$ for the background.
\item $\cos\theta^\prime(\pi) > -0.8$, where $\theta^\prime(\pi)$ is the angle 
between the $\Jpsi$ candidate momentum 
and the pion momentum in the $K^+K^-\pi^+$ rest frame. 
Its distribution is nearly uniform for the
signal processes but peaks towards $-1$ for the background.
\end{itemize}
Distributions of these two variables after applying all other selection 
requirements described in this section are shown in 
Fig.~\ref{fig:aux:BcDs_angles}. They
are shown for the simulated signal samples, as well as for sidebands of the
mass spectrum in data, defined as the regions $5640\MeV < 
m(\Jpsi\Ds) < 5900$~\MeV\ (left sideband) and $6360\MeV < m(\Jpsi\Ds) < 
6760$~\MeV\ (right sideband). A dip in the $\cos\theta^\prime(\pi)$
distribution for the $\Bc\to\Jpsi\Ds$ signal is caused by rejection of 
$\Bs\to\Jpsi\phi$ candidates discussed below.

\begin{figure}[htbp]
\centering
\subfloat[]{\label{fig:aux:BcDs_angles:BcDs_ctsDs}
\includegraphics[width=0.48\textwidth]{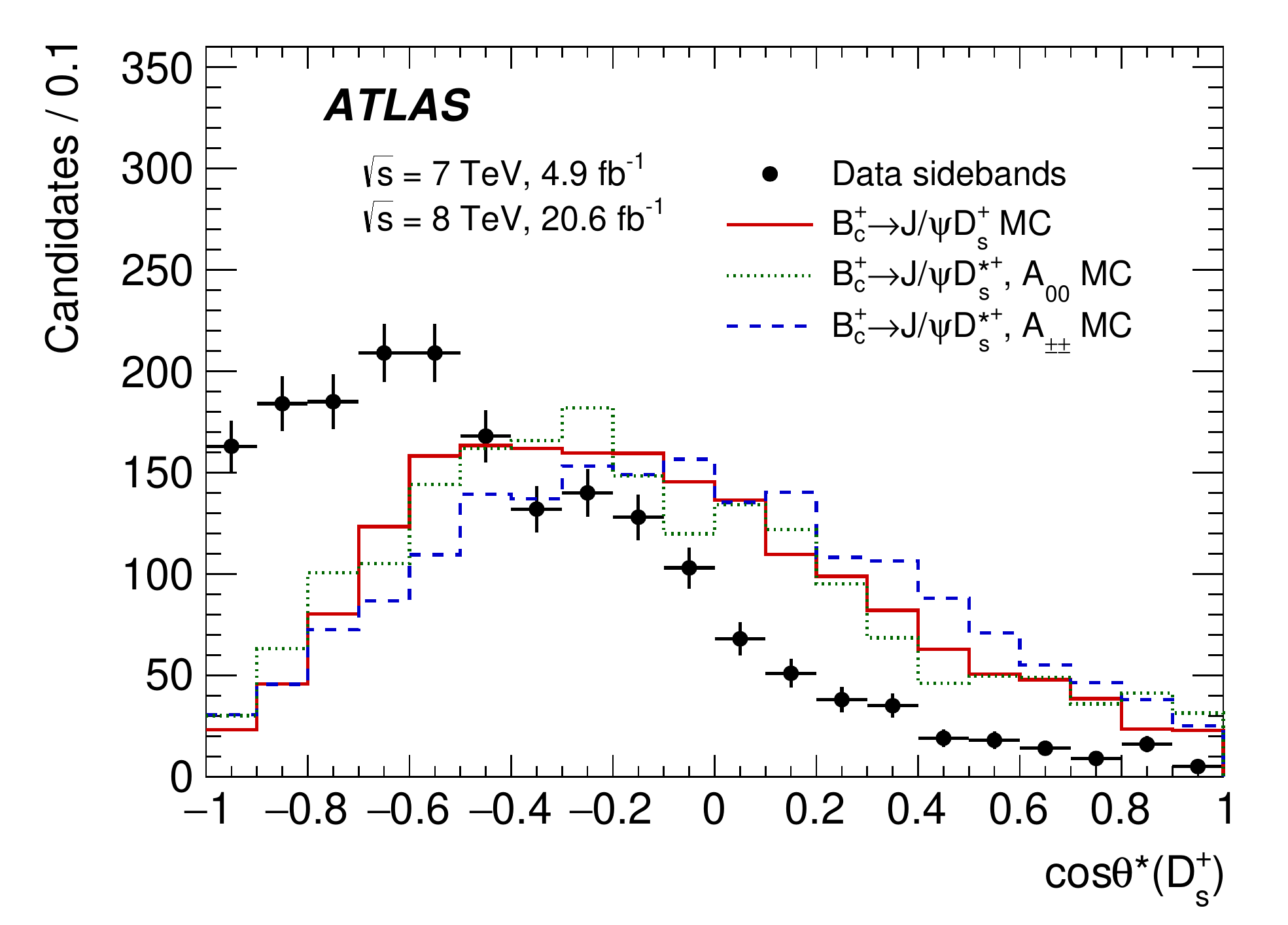}
}
\subfloat[]{\label{fig:aux:BcDs_angles:BcDs_ctpPi}
\includegraphics[width=0.48\textwidth]{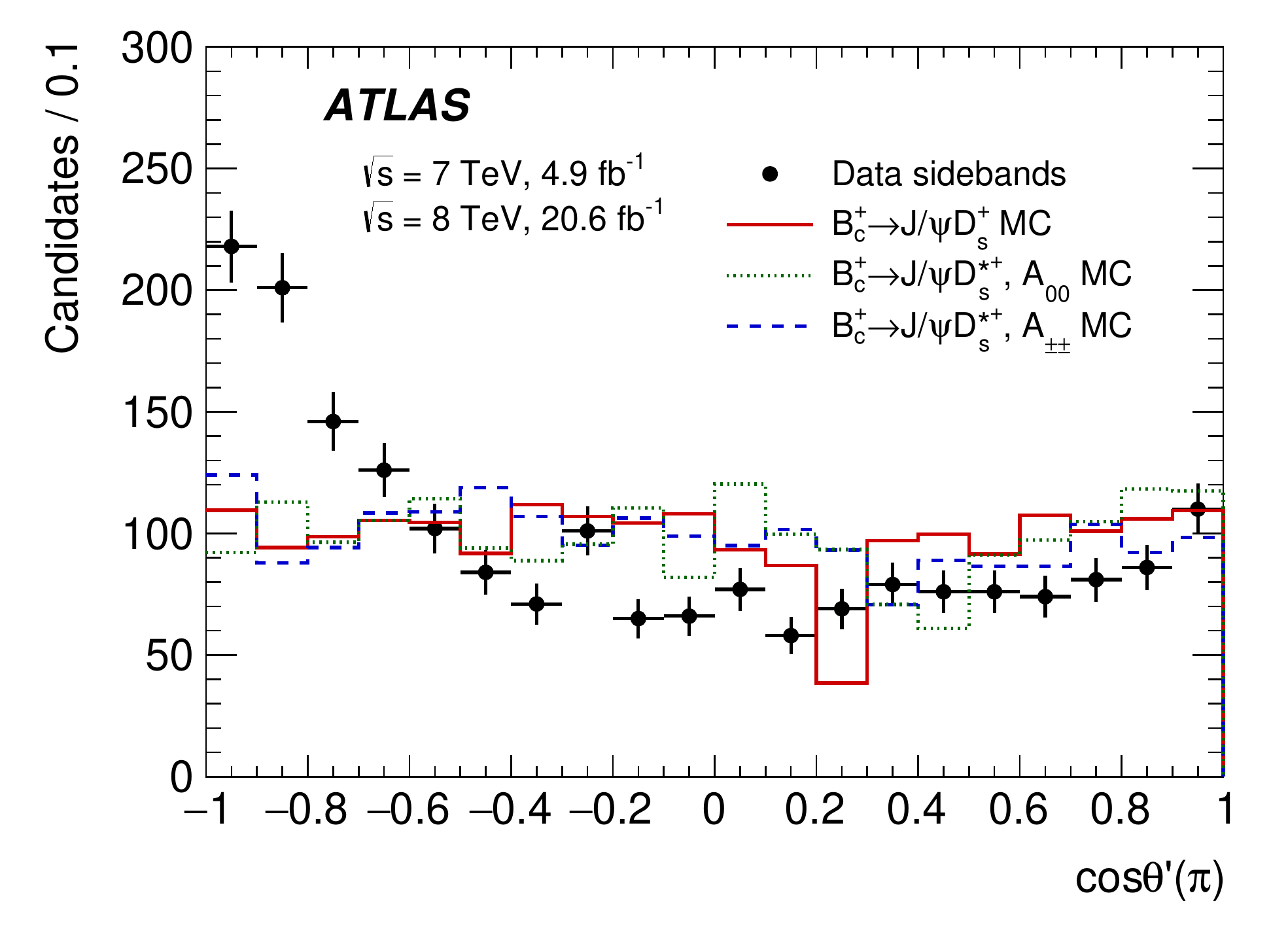}
}
\caption{Distributions of \protect\subref{fig:aux:BcDs_angles:BcDs_ctsDs} 
$\cos\theta^*(\Ds)$ 
and \protect\subref{fig:aux:BcDs_angles:BcDs_ctpPi} $\cos\theta^\prime(\pi)$, 
where $\theta^*(\Ds)$ and $\theta^\prime(\pi)$ are two angular variables 
defined in Sect.~\ref{sec:Reco}. The distributions are shown
for data sidebands (black 
dots) and MC simulation of $\Bc\to\Jpsi\Ds$ signal (red solid line) and 
$A_{00}$ (green dotted line) and $A_{\pm\pm}$ (blue 
dashed line) components of $\Bc\to\Jpsi\DsStar$ signal. The 
distributions are obtained after applying all selection criteria except the 
ones on the plotted variable. The MC distributions are normalised 
to data. 
\label{fig:aux:BcDs_angles}}
\end{figure}

Various possible contributions of partially reconstructed 
$B \to \Jpsi X$ decays were studied. 
The only significant one was found from the $\Bs\to\Jpsi\phi$ decay 
process. This contribution arises when the combination of the tracks from a
true $\Bs\to\Jpsi(\mu^+\mu^-)\phi(K^+K^-)$ decay with a fifth random 
track results in a fake $\Bc\to\Jpsi(\mu^+\mu^-)\Ds(K^+K^-\pi^+)$ 
candidate.
For each reconstructed $\Bc$ candidate, an 
additional vertex fit is performed. The two muon tracks and 
the two kaon tracks are fitted to a common 
vertex, where the kaon tracks are assumed to be from $\phi\to K^+K^-$ and the 
muon pair is constrained to have the nominal $\Jpsi$ mass.
The mass of the $\Bs$ candidate, $m(\mu^+\mu^-K^+K^-)$, is then calculated from 
the refitted track parameters.
Candidates with $5340\MeV < m(\mu^+\mu^-K^+K^-) < 5400$\,\MeV\ are rejected. 
This requirement suppresses the bulk of the $\Bs$ events while rejecting only 
$\sim 4$\% of the signal.

After applying the selection requirements described above, 1547 $\Jpsi\Ds$ 
candidates are selected in the mass range 5640--6760~\MeV.

\section{\texorpdfstring{$\Bc\to\Jpsi\DsOrStar$}{Bc -> J/psi Ds(*)} candidate 
fit}\label{sec:BcDsFit}
The mass distribution of the selected $\Bc\to\Jpsi\DsOrStar$ candidates is 
shown in Fig.~\ref{fig:BcDs_FitMass}. The peak near the $\Bc$ 
mass, $m_{\Bc} = 6275.6$\,\MeV~\cite{Agashe:2014kda}, is attributed to the 
signal of $\Bc\to\Jpsi\Ds$ decay while a wider structure between 5900\,\MeV\ 
and 6200\,\MeV\ corresponds to $\Bc\to\Jpsi\DsStar$ with subsequent 
$\DsStar\to\Ds\gamma$ or 
$\DsStar\to\Ds\pi^0$ decays where the neutral particle is not reconstructed.

\begin{figure*}[htbp]
\centering
\includegraphics[width=0.6\textwidth]{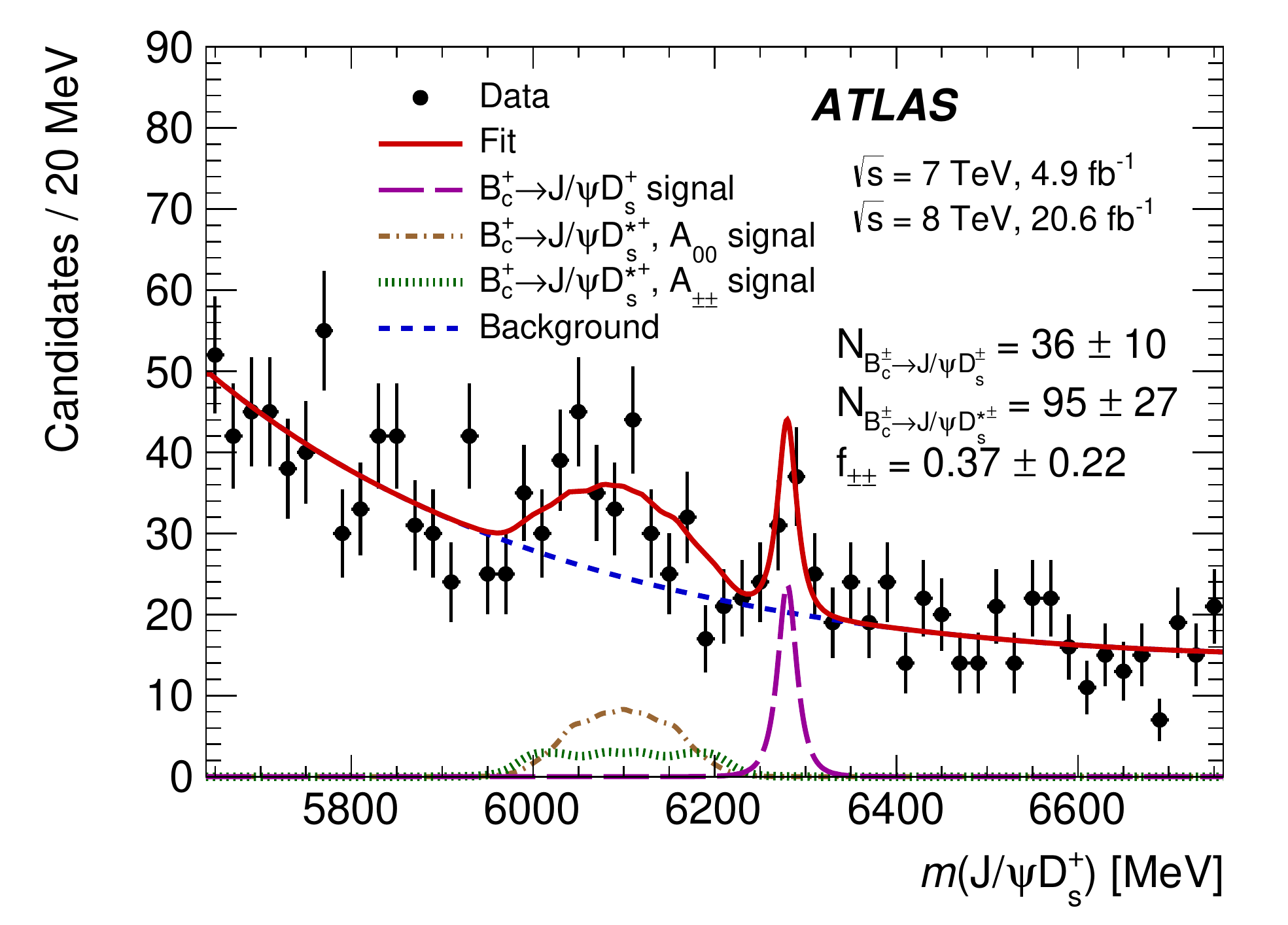}
\caption{The mass distribution for the selected $\Jpsi\Ds$ 
candidates. The red 
solid line represents the projection of 
the fit to the model 
described in the text. The contribution of the $\Bc\to\Jpsi\Ds$ decay is 
shown with the magenta long-dashed line; the brown dash-dot and green dotted 
lines show the 
$\Bc\to\Jpsi\DsStar$ $A_{00}$ and $A_{\pm\pm}$ component contributions, 
respectively; the blue dashed line shows the background model. The 
uncertainties of the listed fit result values are statistical only.
\label{fig:BcDs_FitMass}}
\end{figure*}

Mass distributions of the $\Jpsi$ and $\Ds$ candidates 
corresponding to the $\Jpsi\Ds$ mass region of the observed 
$\Bc\to\Jpsi\DsOrStar$ signals are shown in 
Fig.~\ref{fig:JpsiDsSignals}. To obtain these plots, the 
$\Bc$ candidates are built without the mass constraints in the cascade 
fit, with the mass of the candidate calculated 
as $m(\Jpsi\Ds) = m(\mu^+\mu^-K^+K^-\pi^+) - 
m(\mu^+\mu^-) + m_{\Jpsi} - m(K^+K^-\pi^+) + m_{\Ds}$, where $m_{\Jpsi}$ and 
$m_{\Ds}$ are the nominal masses of the respective particles.
The mass of the $\Bc$ candidate is 
required to be $5900\MeV < m(\Jpsi\Ds) < 6400$\,\MeV\ while the mass windows for the 
corresponding intermediate resonances are widened to the plotting ranges. 
The $\Jpsi$ and $\Ds$ mass distributions are fitted with a sum of an 
exponential function describing the background and a modified 
Gaussian function~\cite{Chekanov:2005cf,Aad:2012jga} describing the corresponding signal 
peak. The modified Gaussian function is defined as
\begin{equation}\label{eq:GausMod}
\mathrm{Gauss}^\mathrm{mod} \sim 
\exp\left(-\frac{x^{1+\frac{1}{1+x/2}}}{2}\right),
\end{equation}
where $x = |m_0 - m|/\sigma$ with the mean mass $m_0$ and width $\sigma$ being 
free parameters.
The fitted masses of $\Jpsi$ ($3095.1\pm 2.4$\,\MeV) and $D_s$ ($1969.0\pm 4.1$\,\MeV) agree with their nominal 
masses, the widths are consistent with those in the 
simulated samples, and the signal yields are found to be $N_{\Jpsi} = 568 \pm 
28$ and $N_{D_s^\pm} = 175 \pm 36$.

\begin{figure*}[htbp]
\centering
\subfloat[]{\label{fig:JpsiDsSignals:Jpsi}
\includegraphics[width=0.48\textwidth]{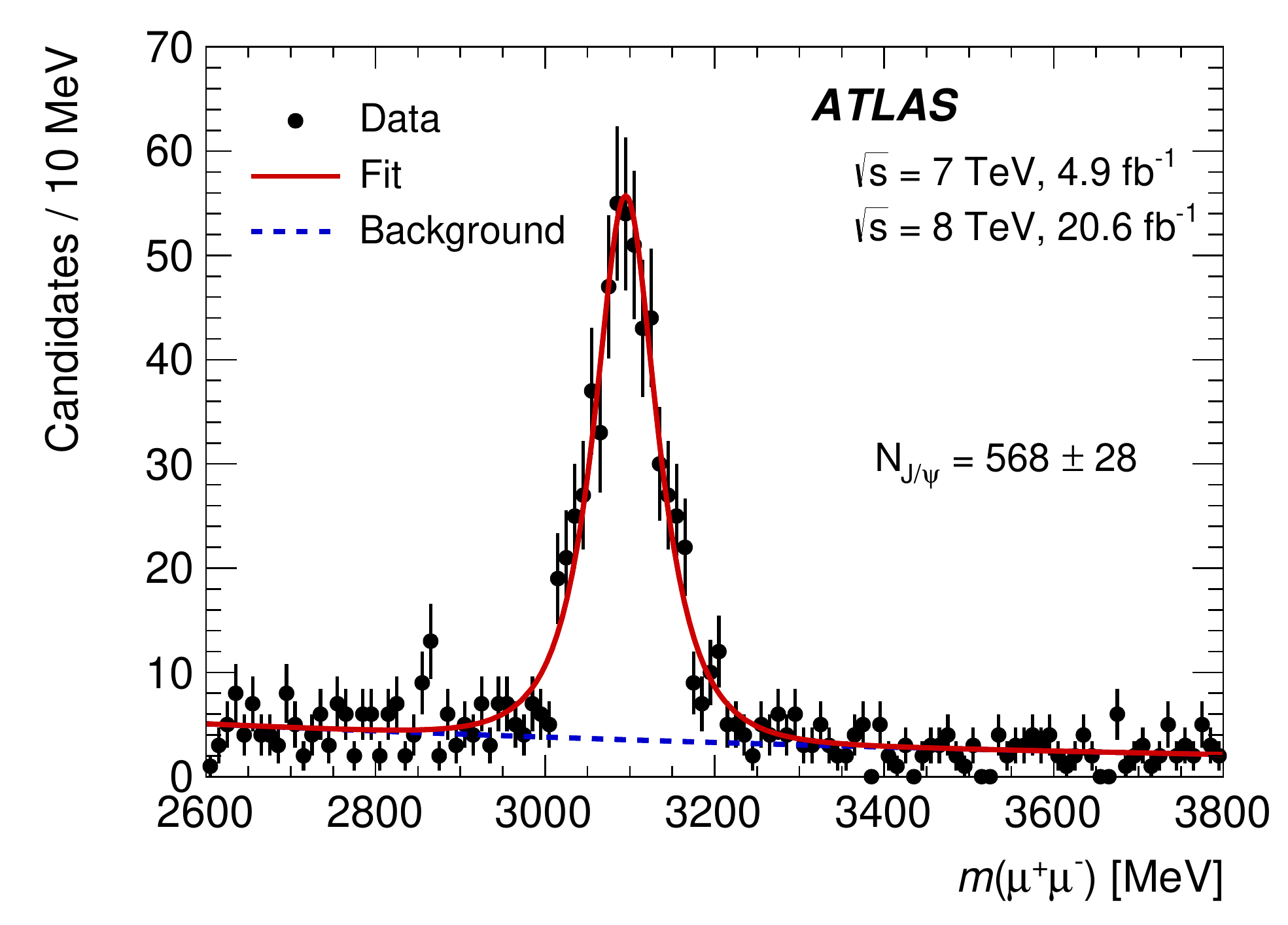}
}
\subfloat[]{\label{fig:JpsiDsSignals:Ds}
\includegraphics[width=0.48\textwidth]{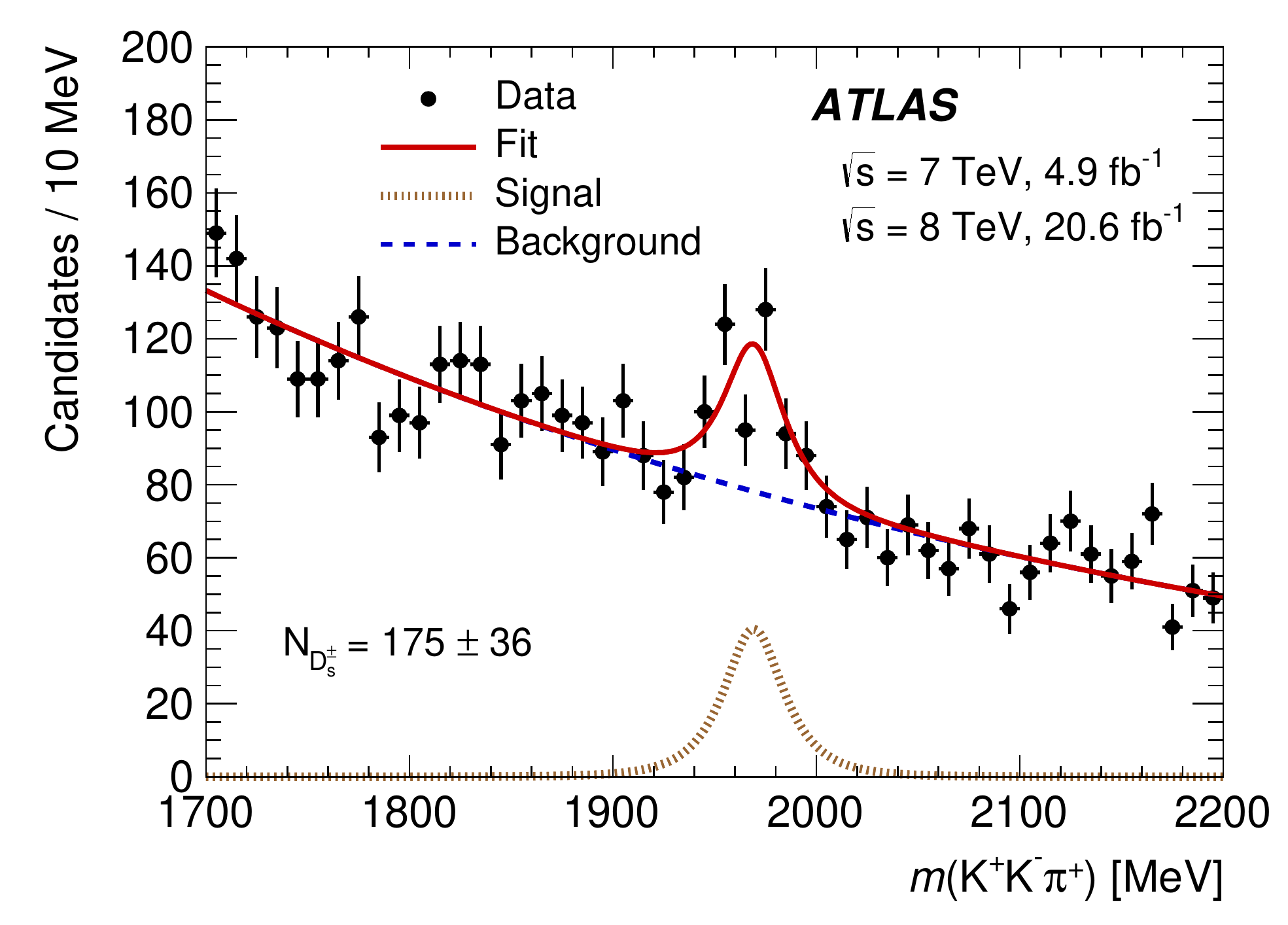}
}
\caption{Mass distribution of the \protect\subref{fig:JpsiDsSignals:Jpsi} 
$\Jpsi$ and \protect\subref{fig:JpsiDsSignals:Ds} $\Ds$ 
candidates after the full 
$\Bc\to\Jpsi\DsOrStar$ selection (without mass constraints in the cascade 
fit) 
in the mass window of the $\Bc$ candidate $5900\MeV < m(\Jpsi\Ds) < 6400$\,\MeV. 
The spectra are fitted with a sum of an exponential and a modified 
Gaussian function. The uncertainties of the shown $\Jpsi$ and $\Ds$ yields are 
statistical only.
\label{fig:JpsiDsSignals}}
\end{figure*}

The information about the helicity in $\Bc\to\Jpsi\DsStar$ decay is encoded 
both in the mass distribution of the $\Jpsi\Ds$ system and in the distribution 
of the helicity angle, $\theta^\prime(\mu^+)$, which is defined in the rest 
frame of the muon pair as the angle between the $\mu^+$ and the $\Ds$ candidate 
momenta.
Thus, a two-dimensional extended unbinned maximum-likelihood fit of the 
$m(\Jpsi\Ds)$ 
and $|\cos\theta^\prime(\mu^+)|$ distributions is performed.
The $A_{++}$ and $A_{--}$ helicity amplitude contributions are described by the 
same mass and angular shapes because of the parity symmetry of the $\Jpsi$ and 
$\DsStar$ decays. This is confirmed by the MC simulation.
Thus these components are treated together as the $A_{\pm\pm}$ component, 
while the shape of the $A_{00}$ component 
is different and is therefore treated separately. A simultaneous fit 
to the 
mass and angular distributions significantly improves the sensitivity to the 
contributions of the helicity amplitudes in $\Bc\to\Jpsi\DsStar$ decay with 
respect to a one-dimensional mass fit.

Four two-dimensional probability density functions (PDFs) are defined to 
describe the $\Bc\to\Jpsi\Ds$ signal, the $A_{\pm\pm}$ and $A_{00}$ components 
of the $\Bc\to\Jpsi\DsStar$ signal, and the background. The signal PDFs are 
factorised into 
mass and angular components. The effect 
of correlations between their mass and angular shapes is found to be 
small and is accounted for as a systematic uncertainty.

The mass distribution of the $\Bc\to\Jpsi\Ds$ signal is described by a modified 
Gaussian function.
For the $\Bc\to\Jpsi\DsStar$ signal components, the mass 
shape templates obtained from the simulation with the kernel estimation 
technique~\cite{Cranmer:2000du} are used. The branching fractions of 
$\DsStar\to\Ds\pi^0$ and $\DsStar\to\Ds\gamma$ decays for the simulation are 
set to the world average values~\cite{Agashe:2014kda}. 
The position of the templates along the 
mass axis is 
varied in the fit simultaneously with the position of the $\Bc\to\Jpsi\Ds$ 
signal peak.
The background mass shape is 
described with a two-parameter exponential function, $\exp\left[a\cdot 
m(\Jpsi\Ds) + b\cdot m(\Jpsi\Ds)^2\right]$.

To describe the $|\cos\theta^\prime(\mu^+)|$ shapes, templates from the kernel 
estimation are used. 
The templates for the signal angular PDFs are extracted from the simulated 
samples. Although their shapes are calculable analytically, using the templates 
allows the fit to account for detector effects.
The background angular description is based on the 
$|\cos\theta^\prime(\mu^+)|$ shape of the candidates in the sidebands of 
$\Jpsi\Ds$ mass spectra. Two templates are produced from the angular 
distributions of the candidates in the left and right mass sidebands as
defined in Sect.~\ref{sec:Reco}. The angular PDF for the background is defined 
as a conditional PDF 
of $|\cos\theta^\prime(\mu^+)|$ given the per-candidate $m(\Jpsi\Ds)$. 
For the 
candidates in the lower half of the left sideband (5640--5770\,\MeV), the 
template from the left sideband is used. Similarly, the template from the right 
sideband is used for the upper half of the right sideband 
(6560--6760\,\MeV). For the candidates in the middle part 
of the mass spectrum (5770--6560\,\MeV), a linear interpolation 
between the two templates is used.

The fit has seven free parameters: the mass of the $\Bc$ meson, 
$m_{\Bc\to\Jpsi\Ds}$; 
the relative contribution of the $A_{\pm\pm}$ component to the total 
$\Bc\to\Jpsi\DsStar$ decay rate in the selected sample, $\fpp$; the two 
parameters of the exponential 
background; the yields of the two signal modes, $N_{\Bc\to\Jpsi\Ds}$ and 
$N_{\Bc\to\Jpsi\DsStar}$, and the background yield.
The width of the modified 
Gaussian function, $\sigma_{\Bc\to\Jpsi\Ds}$, is fixed to the value obtained 
from the 
fit to the simulated signal, $\sigma_{\Bc\to\Jpsi\Ds} = 9.95$\,\MeV. Leaving this 
parameter free in the data fit results in the value $7.9 \pm 3.0$\,\MeV, consistent with the simulation 
in the range of statistical uncertainty.

It was checked that the fit procedure provides unbiased 
values and correct 
statistical uncertainties for the extracted parameters using pseudo-experiments.
The values of the relevant parameters obtained from the fit are given in 
Table~\ref{tab:BcDsFitResults}. The fitted 
$\Bc$ mass agrees with the world average value~\cite{Agashe:2014kda}.
The mass 
and angular projections of the fit on the selected $\Jpsi\Ds$ candidate dataset 
are also shown in 
Figs.~\ref{fig:BcDs_FitMass} 
and~\ref{fig:BcDs_FitAng}\subref{fig:BcDs_FitAng:Full}, respectively. In 
order to illustrate the effect of the angular part of the fit in separating the 
helicity 
amplitudes, the $|\cos\theta^\prime(\mu^+)|$ projection for the subset 
of candidates with the masses $5950\MeV < m(\Jpsi\Ds) < 6250$\,\MeV\ 
corresponding to the region 
of the observed $\Bc\to\Jpsi\DsStar$ signal is shown in 
Fig.~\ref{fig:BcDs_FitAng}\subref{fig:BcDs_FitAng:DsStarSlice}.

\begin{table}[htbp]
\begin{center}
\caption{Parameters of the $\Bc\to\Jpsi\DsOrStar$ signals obtained with the 
unbinned extended maximum-likelihood fit. The width parameter of the 
modified Gaussian function is fixed to the MC value. Only statistical 
uncertainties are 
shown. No acceptance corrections are applied to the signal yields.
\label{tab:BcDsFitResults}}
\vspace{1em}
\begin{tabular}{l r @{$\;\pm\;$} l}
\hline\hline
Parameter  & \multicolumn{2}{r}{Value}   \\
\hline
$m_{\Bc\to\Jpsi\Ds}$ [\MeV]               & $6279.9$ & $3.5$ \\
$N_{\Bc\to\Jpsi\Ds}$          & $36$ & $10$ \\
$N_{\Bc\to\Jpsi\DsStar}$      & $95$ & $27$ \\
$\fpp$                    & $0.37$ & $0.22$ \\
\hline\hline
\end{tabular}
\end{center}
\end{table}

\begin{figure*}[htbp]
\centering
\subfloat[]{
\includegraphics[width=0.48\textwidth]{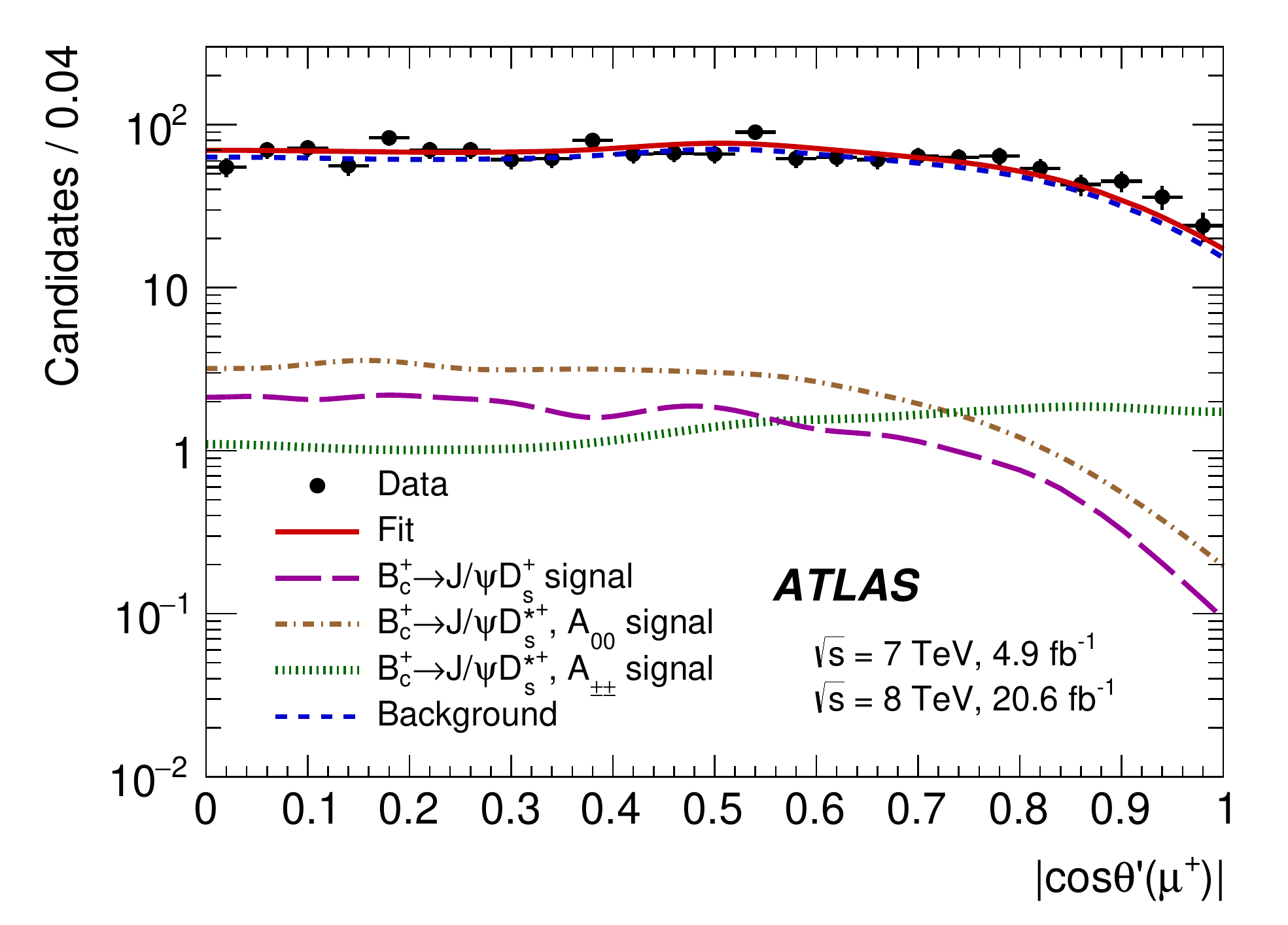}
\label{fig:BcDs_FitAng:Full}
}
\subfloat[]{
\includegraphics[width=0.48\textwidth]{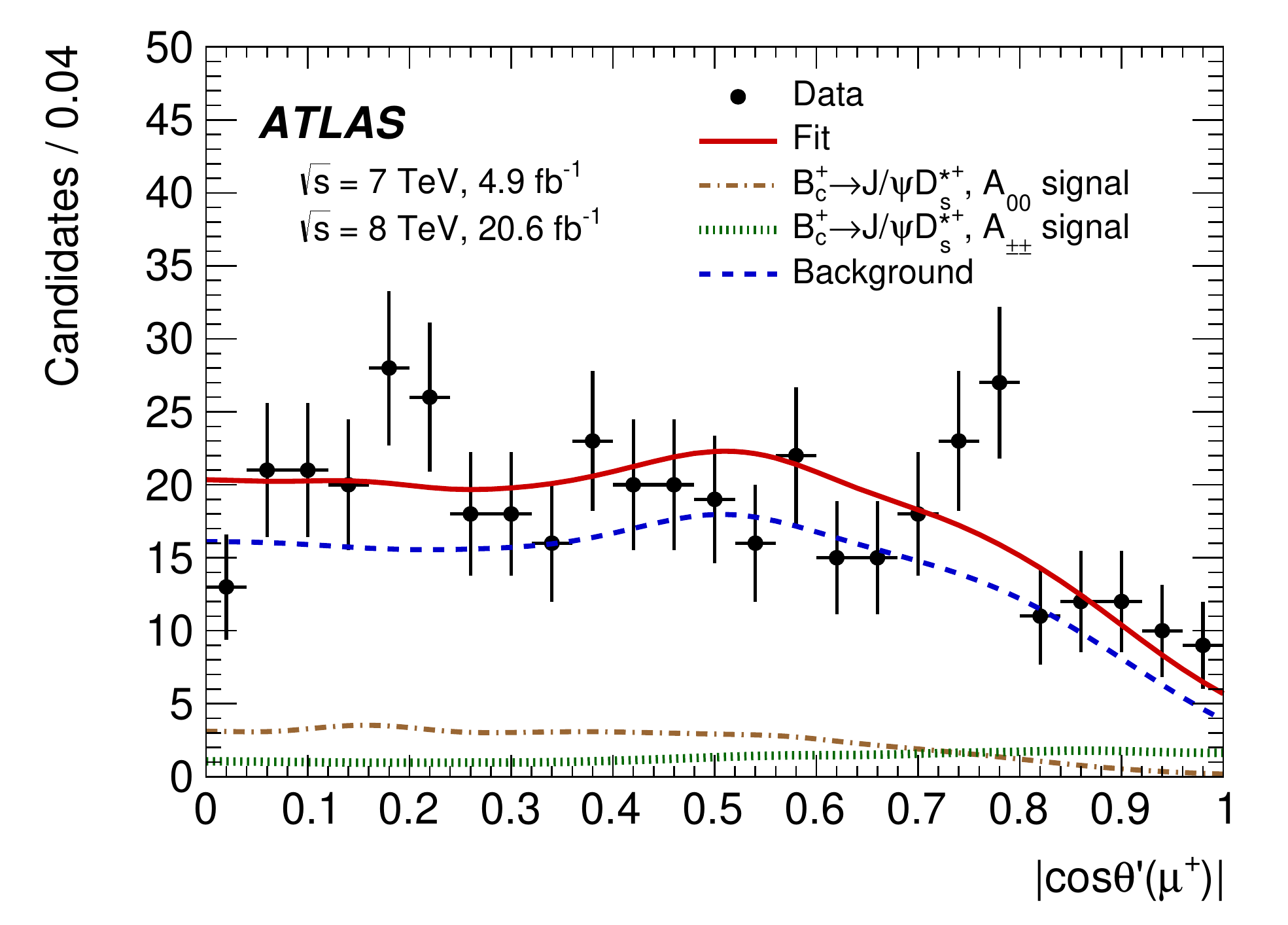}
\label{fig:BcDs_FitAng:DsStarSlice}
}
\caption{The projection of the likelihood fit on the variable 
$|\cos\theta^\prime(\mu^+)|$, where the helicity angle $\theta^\prime(\mu^+)$ 
is the angle between the $\mu^+$ and $\Ds$ candidate momenta in the rest frame 
of the muon pair from $\Jpsi$ decay, for
\protect\subref{fig:BcDs_FitAng:Full} the full selected 
$\Jpsi\Ds$ candidate dataset and \protect\subref{fig:BcDs_FitAng:DsStarSlice} 
a subset of the candidates in a mass range $5950\MeV < m(\Jpsi\Ds) < 
6250$\,\MeV\ corresponding to the observed 
signal of $\Bc\to\Jpsi\DsStar$ decay. The red 
solid line represents the full fit projection. The 
contribution of the $\Bc\to\Jpsi\Ds$ decay is 
shown with the magenta long-dashed line (it is not drawn in 
\protect\subref{fig:BcDs_FitAng:DsStarSlice} because this contribution vanishes 
in that mass range); the brown dash-dot and green dotted 
lines show the 
$\Bc\to\Jpsi\DsStar$ $A_{00}$ and $A_{\pm\pm}$ component contributions, 
respectively; the blue dashed line shows the background model.
\label{fig:BcDs_FitAng}}
\end{figure*}

The statistical significance for the observed $\Bc$ signal estimated from 
toy MC studies is 4.9 standard deviations.

\section{\texorpdfstring{$\Bc\to\Jpsi\pi^+$}{Bc -> J/psi pi} candidate 
reconstruction and fit}\label{sec:BcPi}
$\Bc\to\Jpsi\pi^+$ candidates are reconstructed by fitting a common vertex of a 
pion candidate 
track and the two muons from a $\Jpsi$ candidate, selected 
as described in Sect.~\ref{sec:Reco}. For the pion candidate, 
tracks identified as muons are vetoed in order to suppress the substantial 
background from $\Bc\to\Jpsi\mu^+\nu_\mu X$ decays. The invariant mass of the 
two muons in the vertex fit is constrained to the $\Jpsi$ nominal mass. 
The quality of the fit must satisfy $\chi^2/\ndf < 3$.
The following selection requirements applied to the $\Bc\to\Jpsi\pi^+$ candidates 
are analogous to those for $\Bc\to\Jpsi\Ds$ candidates described in Sect.~\ref{sec:Reco}: 
the candidates must be within the kinematic range $\pT(\Bc) > 15$\,\GeV, $|\eta(\Bc)| < 2.0$; 
the refitted values of the transverse momenta and pseudorapidities of the muons are required to satisfy $\pT(\mu^\pm) > 3$\,\GeV,
$|\eta(\mu^\pm)|<2.3$; the same requirements on pointing to the primary vertex and the ratio 
$\pT(\Bc)/\sum\pT(\mathrm{trk})$ are applied. The refitted pion track kinematics must satisfy $\pT(\pi^+) > 5$\,\GeV\ 
and $|\eta(\pi^+)|<2.5$.
The transverse decay length is required to be $L_{xy}(\Bc) > 0.2$~mm, and not to 
exceed 10~mm. 

%

To further suppress combinatorial background, the following selection is 
applied:
\begin{itemize}
\item $\cos\theta^*(\pi) > -0.8$, where $\theta^*(\pi)$ is the angle 
between the pion momentum in the $\mu^+\mu^-\pi^+$ rest frame and the $\Bc$ 
candidate line of flight in laboratory frame. This angular variable behaviour 
for the signal and the background is the same as that of $\cos\theta^*(\Ds)$ 
used for $\Jpsi\Ds$ candidates selection.
\item $|\cos\theta^\prime(\mu^+)| < 0.8$, where $\theta^\prime(\mu^+)$ is the 
angle between the $\mu^+$ and $\pi^+$ momenta in the muon pair rest frame. The 
signal distribution follows a $\sin^2\theta^\prime(\mu^+)$ shape, while the 
background is flat.
\end{itemize}

After applying the above-mentioned requirements, 38542 $\Jpsi\pi^+$ 
candidates are selected in the mass range 5640--6760~\MeV. 
Figure~\ref{fig:BcPi_MassFit} shows 
the mass distribution of the selected candidates. 
An extended unbinned maximum-likelihood fit of the mass 
spectrum is performed to evaluate the $\Bc\to\Jpsi\pi^+$ signal yield. The 
signal contribution is described with the modified Gaussian function while an 
exponential function is used for the background. The $\Bc$ 
mass, $m_{\Bc\to\Jpsi\pi^+}$, the width of the modified 
Gaussian function, 
$\sigma_{\Bc\to\Jpsi\pi^+}$, the yields of the signal, $N_{\Bc\to\Jpsi\pi^+}$, 
and the background, and the slope of the exponential background are free 
parameters of the fit. 
The fit results are summarised in Table~\ref{tab:BcPiFitResults}, and the 
fit projection is also shown in 
Fig.~\ref{fig:BcPi_MassFit}. The extracted $\Bc$ mass value is consistent with 
the world average~\cite{Agashe:2014kda}, and the signal peak width agrees 
with the simulation (37.4~\MeV).

\begin{table}[htbp]
\begin{center}
\caption{Signal parameters of the $\Jpsi\pi^+$ mass 
distribution obtained with the unbinned extended maximum-likelihood 
fit. Only statistical uncertainties are shown. No acceptance corrections are 
applied to the signal yields. \label{tab:BcPiFitResults}}
\vspace{1em}
\begin{tabular}{l r @{$\;\pm\;$} l}
\hline\hline
Parameter  & \multicolumn{2}{r}{Value}   \\
\hline
$m_{\Bc\to\Jpsi\pi^+}$ [\MeV]               & $6279.9$ & $3.9$ \\
$\sigma_{\Bc\to\Jpsi\pi^+}$ [\MeV]          & $33.9$ & $4.2$ \\
$N_{\Bc\to\Jpsi\pi^+}$                      & $1140$ & $120$ \\
\hline\hline
\end{tabular}
\end{center}
\end{table}

\begin{figure}[htbp]
\centering
\includegraphics[width=0.48\textwidth]{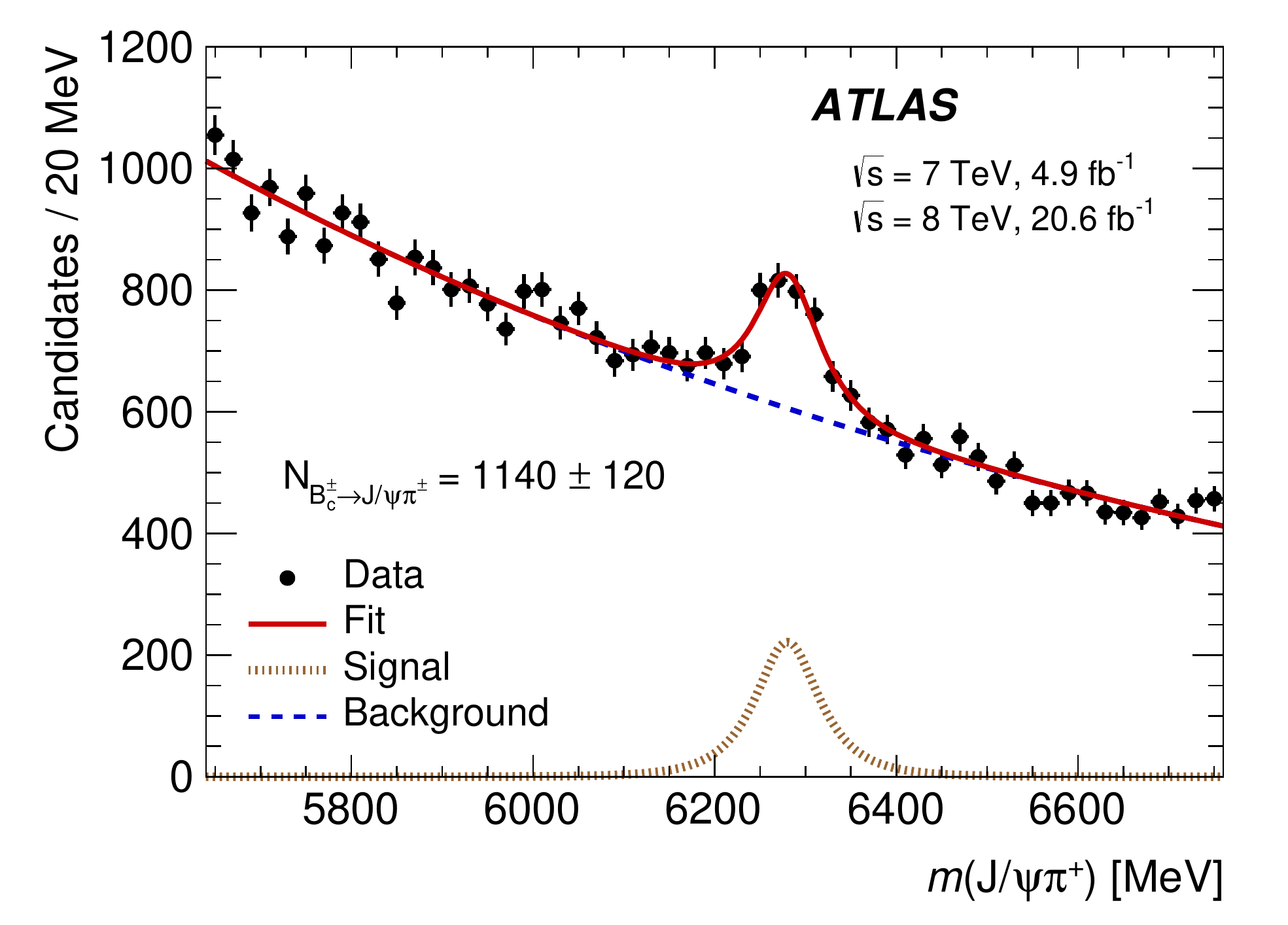}
\caption{The mass distribution for the selected $\Bc\to\Jpsi\pi^+$ 
candidates. The red solid line represents the result of the fit to the model 
described in the text. The brown dotted and blue dashed lines show the 
signal and background component projections, respectively. The uncertainty of 
the shown signal yield is statistical only.
\label{fig:BcPi_MassFit}}
\end{figure}

\section{Branching fractions and polarisation measurement}\label{sec:RelBr}
The ratios of the branching fractions $\R_{\Ds/\pi^+}$ and $\R_{\DsStar/\pi^+}$
are calculated as
\begin{equation}\label{eq:RelBrDs}
\R_{\DsOrStar/\pi^+} = 
\frac{\Br_{\Bc\to\Jpsi\DsOrStar}}{\Br_{\Bc\to\Jpsi\pi^+}} = 
\frac{1}{\Br_{\Ds\to\phi(K^+K^-)\pi^+}} \times 
\frac{\A_{\Bc\to\Jpsi\pi^+}}{\A_{\Bc\to\Jpsi\DsOrStar}} \times 
\frac{N_{\Bc\to\Jpsi\DsOrStar}}{N_{\Bc\to\Jpsi\pi^+}},
\end{equation}
where $\A_{\Bc\to X}$ and 
$N_{\Bc\to X}$ are the total acceptance and 
the yield of the corresponding mode.
For $\Br_{\Ds\to\phi(K^+K^-)\pi^+}$, the CLEO 
measurement~\cite{Alexander:2008aa} of the partial $\Ds\to K^+K^-\pi^+$ 
branching fractions, with a kaon-pair mass within various intervals around the 
nominal $\phi$ meson mass, is used.
An interpolation between the partial branching 
fractions, measured for $\pm 5$\,\MeV\ and $\pm 10$\,\MeV\ intervals, using a
relativistic Breit--Wigner shape of the resonance yields 
the value $(1.85 \pm 0.11)$\% for the $\pm 7$\,\MeV\ interval which is used 
in the analysis. 
The effect of admixture of other $\Ds$ decay modes with $(K^+K^-\pi^+)$ final state which are not present in the MC simulation
is studied separately and accounted for as a systematic uncertainty.

The acceptance for the $\Bc\to\Jpsi\DsStar$ decay mode 
is different
for the $A_{\pm\pm}$ and $A_{00}$ components, thus the full 
acceptance for the mode is
\begin{equation}\label{eq:AccDsStar}
\A_{\Bc\to\Jpsi\DsStar} = 
\left(\frac{\fpp}{\A_{\Bc\to\Jpsi\DsStar,A_{\pm\pm}}} + 
\frac{1-\fpp}{\A_{\Bc\to\Jpsi\DsStar,A_{00}}}\right)^{-1},
\end{equation}
where the subscripts indicate the helicity state and $\fpp$ is the value 
extracted 
from the fit (Table~\ref{tab:BcDsFitResults}).
The acceptances are determined from the simulation and shown in 
Table~\ref{tab:acceptances}.

\begin{table}[htbp]
\begin{center}
\caption{The acceptance $\A_{\Bc\to X}$ for 
all decay modes studied. Only uncertainties due to MC statistics are 
shown.
\label{tab:acceptances}}
\vspace{1em}
\begin{tabular}{l r}
\hline\hline
Mode & $\A_{\Bc\to X}$ [\%] \\
\hline
$\Bc\to\Jpsi\pi^+$                 & $4.106 \pm 0.056$ \\
$\Bc\to\Jpsi\Ds$                   & $1.849 \pm 0.034$ \\
$\Bc\to\Jpsi\DsStar$, $A_{00}$     & $1.829 \pm 0.053$ \\
$\Bc\to\Jpsi\DsStar$, $A_{\pm\pm}$ & $1.712 \pm 0.035$ \\
\hline\hline
\end{tabular}
\end{center}
\end{table}

The ratio $\R_{\DsStar/\Ds}$ is 
calculated as
\begin{equation}\label{eq:RelBrDsDsStar}
\R_{\DsStar/\Ds} = \frac{\Br_{\Bc\to\Jpsi\DsStar}}{\Br_{\Bc\to\Jpsi\Ds}} = 
\frac{N_{\Bc\to\Jpsi\DsStar}}{N_{\Bc\to\Jpsi\Ds}}  \times 
\frac{\A_{\Bc\to\Jpsi\Ds}}{\A_{\Bc\to\Jpsi\DsStar}},
\end{equation}
where the ratio of the yields 
${N_{\Bc\to\Jpsi\DsStar}}/{N_{\Bc\to\Jpsi\Ds}}$ and its uncertainty is 
extracted from the fit as a parameter 
in order to account for correlations between the yields. 

The fraction of the $A_{\pm\pm}$ component contribution in 
$\Bc\to\Jpsi\DsStar$ decay 
is calculated from the $\fpp$ value quoted in 
Table~\ref{tab:BcDsFitResults} by applying a correction to account for 
the different acceptances for the two component contributions:
\begin{equation}\label{eq:fPlusPlusCor}
\Gamma_{\pm\pm}/\Gamma = 
\fpp \times
\frac{\A_{\Bc\to\Jpsi\DsStar}}{\A_{\Bc\to\Jpsi\DsStar,\A_{\pm\pm}}}.
\end{equation}

\section{Systematic uncertainties}\label{sec:Syst}
The systematic uncertainties  of the measured values are determined by varying 
the analysis procedure and repeating all calculations. Although some sources 
can have rather large effects on the individual decay rate measurements, 
they largely cancel in the ratios of the branching fractions due to 
correlation between the effects on the different decay modes. The following 
groups of systematic uncertainties are considered.

The first group of sources of systematic uncertainty relates to possible 
differences between 
the data and simulation affecting the acceptances for the decay modes.
Thus, an effect of the $\Bc$ production model is evaluated by varying the 
simulated 
$\pT$ and $|\eta|$ spectra while preserving agreement with the data 
distributions obtained using the abundant $\Bc\to\Jpsi\pi^+$ channel.
These variations have very similar effects on the acceptances for the different 
decay modes, thus giving rather moderate estimates of the uncertainties, not 
exceeding 3\% in total, on the ratios of branching fractions.
The effect of presence of other $\Ds$ decay modes with $(K^+K^-\pi^+)$ final state 
on the calculated acceptances is studied with a separate MC simulation. Its conservative estimate 
yields 0.4\% which is assigned as $\R_{\Ds/\pi^+}$ and $\R_{\DsStar/\pi^+}$ uncertainties.
An uncertainty on the 
tracking efficiency is 
dominated by the uncertainty of the detector material description in the MC 
simulation. Samples generated
with distorted geometries and with increased material are used to estimate the 
effect on track reconstruction efficiencies. 
When propagated to the ratios of branching fractions, these estimates give 0.5\% 
uncertainty for 
$\R_{\Ds/\pi^+}$ and $\R_{\DsStar/\pi^+}$ due to the two extra tracks in 
$\Bc\to\Jpsi\DsOrStar$ modes.
Limited knowledge of the $\Bc$ and $\Ds$ lifetimes leads to an additional 
systematic uncertainty. The simulated proper decay times are varied within one 
standard deviation from the world average 
values~\cite{Agashe:2014kda} resulting in uncertainties of $\sim 1$\% 
assigned to $\R_{\Ds/\pi^+}$ and $\R_{\DsStar/\pi^+}$ due to the $\Bc$ 
lifetime, and 0.3\% due to the $\Ds$ lifetime. 
Removing the requirement on 
$\pT(\Bc)/\sum\pT(\mathrm{trk})$ is found to produce no noticeable effect 
on the measured values.

The next group of uncertainties originates from the signal extraction 
procedure. 
These uncertainties are evaluated separately for $\Jpsi\Ds$ and $\Jpsi\pi^+$ 
candidate fits. 
For the former, the following variations of the fit model are applied 
and the difference is treated as a systematic uncertainty:

\begin{itemize}
\item different background mass shape parametrisations (three-parameter 
exponential, second- and third-order 
polynomials), different fitted mass range 
(reduced by up to 40\,\MeV\ from each side independently);

\item a double Gaussian or double-sided Crystal Ball 
function~\cite{Oreglia:1980cs, Gaiser:1982yw, Skwarnicki:1986xj} 
for 
$\Bc\to\Jpsi\Ds$ signal 
description; variation of the modified Gaussian width within 10\% of the MC 
simulation value;

\item variation of the smoothness of the $\Bc\to\Jpsi\DsStar$ signal mass 
templates, which is controlled by a parameter of the kernel estimation 
procedure~\cite{Cranmer:2000du};

\item similar variation of the smoothness of the $\Bc\to\Jpsi\DsOrStar$ signal 
angular templates;

\item variation of the smoothness of the sideband templates used for the 
background angular PDF construction; different ranges of the sidebands; 
different sideband interpolation procedure;

\item modelling of the correlation between the mass and 
angular parts of the signal PDFs. This correlation takes place only at the 
detector level and manifests itself in degradation of the mass resolution 
for higher values of $|\cos\theta^\prime(\mu^+)|$. A dedicated fit model 
accounting for this effect is used for the data fit. The impact on the 
result is found to be negligible compared to the total uncertainty.

\end{itemize}
The first two items give the dominant contributions to the uncertainties of the 
ratios of branching fractions while the transverse polarisation fraction 
measurement is mostly affected by the background angular modelling variations.
For the normalisation channel fit model, the similar variations of the 
background and signal mass shape parametrisation are applied.
The deviations produced by the variations of the fits reach values as high as 
10--15\% thus making them the dominant sources of systematic uncertainty.

The branching fractions of $\DsStar$~\cite{Agashe:2014kda} are varied in 
simulation within their uncertainties to estimate their effect on the measured 
quantities. Very small uncertainties are obtained for the  $\R_{\DsStar/\pi^+}$ 
and $\R_{\DsStar/\Ds}$, while for $\Gamma_{\pm\pm}/\Gamma$, the estimate is 
$\sim 1$\%.

The statistical uncertainties on the acceptance values due to the MC 
sample sizes are also treated as a separate source of systematic uncertainty 
and 
estimated to be 2--3\%.

In order to check for a possible bias from using three-muon 
triggers, vetoing the $\Ds$ meson daughter tracks identified as muons is 
tested and found not to affect the measurement.

Finally, since $\Br_{\Ds\to\phi(K^+K^-)\pi^+}$ enters 
Eq.~(\ref{eq:RelBrDs}), its uncertainty, evaluated from 
Ref.~\cite{Alexander:2008aa} as 5.9\%,
is propagated to the final values of the relative branching fractions.

The systematic uncertainties on the measured quantities are summarised in 
Table~\ref{tab:uncertainties}.

\begin{table*}[htbp]
\begin{center}
\caption{Relative systematic uncertainties on the measured ratios
of branching fractions $R_{\Ds/\pi^+}$, $R_{\DsStar/\pi^+}$, $R_{\DsStar/\Ds}$ 
and on the transverse polarisation fraction $\Gamma_{\pm\pm}/\Gamma$.
\label{tab:uncertainties}}
\begin{tabular}{l r r r r}
\hline\hline
Source & \multicolumn{4}{c}{Uncertainty [\%]}    \\
       & $R_{\Ds/\pi^+}$ 
& $R_{\DsStar/\pi^+}$ & 
$R_{\DsStar/\Ds}$ & $\Gamma_{\pm\pm}/\Gamma$ \\[0.2em]
\hline\\[-0.9em]

Simulated $\pT(\Bc)$ spectrum & $0.4$ & $0.9$ & 
$0.5$  & $0.4$\\[0.4em]

Simulated $|\eta(\Bc)|$ spectrum & $1.9$ & $2.4$ & 
$0.6$ & $0.2$\\[0.4em]

Other $\Ds$ decay modes contribution & $0.4$ & $0.4$ & 
-- & --\\[0.4em]

Tracking efficiency & $0.5$ & $0.5$ & $<0.1$ & $<0.1$ 
\\[0.4em]

$\Bc$ lifetime & $1.2$ & $1.3$ & $<0.1$ & $<0.1$ \\[0.4em]

$\Ds$ lifetime & $ 0.3$ & $ 0.3$ &  
$<0.1$ & $<0.1$ \\[0.4em]

$\Bc\to\Jpsi\DsOrStar$ signal extraction & $4.4$ & 
$10.5$ & $10.7$ & $17.4$ \\[0.4em]

$\Bc\to\Jpsi\pi^+$ signal extraction & $8.5$ & $8.5$ & -- 
& -- \\[0.4em]

$\DsStar$ branching fractions & $<0.1$ & $<0.1$ & 
$<0.1$ & $1.1$ \\[0.4em]

MC sample sizes & $2.3$ & $2.4$ & 
$2.7$ & $2.2$ \\[0.2em]
\hline\\[-0.9em]

Total & $10.1$ & $14.0$ & 
$11.0$ & 
$17.6$ \\[0.2em]
\hline\\[-0.9em]

$\Br_{\Ds\to\phi(K^+K^-)\pi^+}$ & $5.9$ & $5.9$ 
& -- & -- \\[0.2em]
\hline\hline

\end{tabular}
\end{center}
\end{table*}

\section{Results}\label{sec:Summary}
The following ratios of the 
branching fractions are measured:
\begin{align}\label{eq:RelBrDsNumber}
\R_{\Ds/\pi^+} = 
\frac{\Br_{\Bc\to\Jpsi\Ds}}{\Br_{\Bc\to\Jpsi\pi^+}} &=
3.8 \pm 1.1\stat \pm 0.4\syst \pm 0.2\bran,\\
\R_{\DsStar/\pi^+} = 
\frac{\Br_{\Bc\to\Jpsi\DsStar}}{\Br_{\Bc\to\Jpsi\pi^+}} &=
10.4 \pm 3.1\stat \pm 1.5\syst \pm 0.6\bran,\\
\R_{\DsStar/\Ds} = \frac{\Br_{\Bc\to\Jpsi\DsStar}}{\Br_{\Bc\to\Jpsi\Ds}} &=
2.8^{+1.2}_{-0.8}\stat \pm 0.3\syst,
\end{align}
where the BF
uncertainty corresponds to the knowledge of 
$\Br_{\Ds\to\phi(K^+K^-)\pi^+}$. The relative 
contribution of the $A_{\pm\pm}$ component in $\Bc\to\Jpsi\DsStar$ 
decay is measured to be
\begin{equation}\label{eq:FpmpmNumber}
\Gamma_{\pm\pm}/\Gamma = 
0.38 \pm 0.23\stat \pm 0.07\syst
\end{equation}

These results are compared with those of the LHCb 
measurement~\cite{Aaij:2013gia} 
and to the expectations from various theoretical calculations in 
Table~\ref{tab:PredicRelBr} and Fig.~\ref{fig:aux:ValComparison}.
The measurement agrees with the LHCb result.
All ratios are well described by the recent perturbative QCD 
predictions~\cite{Rui:2014tpa}. 
The expectations from models in Refs.~\cite{Colangelo:1999zn, Ivanov:2006ni, 
Ke:2013yka} as well as the sum-rules prediction~\cite{Kiselev:2002vz} for the 
ratio $\R_{\DsStar/\Ds}$ are consistent with the measurement.
The QCD relativistic potential model predictions~\cite{Colangelo:1999zn} are 
consistent with the measured $\R_{\Ds/\pi^+}$ ratio while the 
expectations from the sum rules~\cite{Kiselev:2002vz} and 
models in Refs.~\cite{Ivanov:2006ni, Dhir:2008hh, Ke:2013yka} are somewhat 
smaller than the measured value.
The predictions in Refs.~\cite{Colangelo:1999zn, Kiselev:2002vz, 
Ivanov:2006ni, Ke:2013yka} are also generally smaller than the measured ratio  
$\R_{\DsStar/\pi^+}$; however, the discrepancies do not exceed two 
standard deviations when taking into account only the experimental uncertainty.

\begin{table}[htbp]
\begin{center}
\caption{
Comparison of the results of this measurement with those of 
LHCb~\cite{Aaij:2013gia} 
and theoretical predictions based on 
a QCD relativistic potential model~\cite{Colangelo:1999zn}, QCD sum 
rules~\cite{Kiselev:2002vz}, relativistic constituent quark model 
(RCQM)~\cite{Ivanov:2006ni}, BSW relativistic quark model 
(with fixed average transverse quark momentum 
$\omega=0.40\GeV$)~\cite{Dhir:2008hh}, 
light-front quark model (LFQM)~\cite{Ke:2013yka}, perturbative QCD 
(pQCD)~\cite{Rui:2014tpa}, and relativistic independent quark model 
(RIQM)~\cite{Kar:2013fna}. The uncertainties of the theoretical predictions are 
shown if they are explicitly quoted in 
the corresponding papers. Statistical and systematic uncertainties added in 
quadrature are shown for the results of ATLAS and LHCb.
\label{tab:PredicRelBr}
}
\vspace{1em}
\begin{tabular}{c c c c r}
\hline\hline
$\R_{\Ds/\pi^+}$  & $\R_{\DsStar/\pi^+}$  & $\R_{\DsStar/\Ds}$  & 
$\Gamma_{\pm\pm}/\Gamma$  & Ref. \\
\hline\\[-0.9em]
$3.8 \pm 1.2$ & $10.4 \pm 3.5$ & $2.8^{+1.2}_{-0.9}$ & $0.38 \pm 
0.24$ &
ATLAS \\[0.2em]
$2.90\pm 0.62$ & -- & $2.37\pm 0.57$ & $0.52 \pm 0.20$ &
LHCb~\cite{Aaij:2013gia} \\
\hline
$2.6$ & $4.5$ & $1.7$ & -- & QCD potential model \cite{Colangelo:1999zn} \\
$1.3$ & $5.2$ & $3.9$ & -- & QCD sum rules \cite{Kiselev:2002vz} \\
$2.0$ & $5.7$ & $2.9$ & -- & RCQM \cite{Ivanov:2006ni} \\
$2.2$ & -- & -- & -- & BSW \cite{Dhir:2008hh} \\
$2.06\pm 0.86$ & -- & $3.01\pm 1.23$ & -- & LFQM \cite{Ke:2013yka} \\[0.2em]
$3.45^{+0.49}_{-0.17}$ & -- & $2.54^{+0.07}_{-0.21}$ & $0.48\pm 0.04$ & 
pQCD \cite{Rui:2014tpa} \\[0.2em]
-- & -- & -- & $0.410$ & RIQM \cite{Kar:2013fna} 
\\
\hline\hline
\end{tabular}
\end{center}
\end{table}

\begin{figure*}[htbp]
\centering
\includegraphics[width=0.96\textwidth]{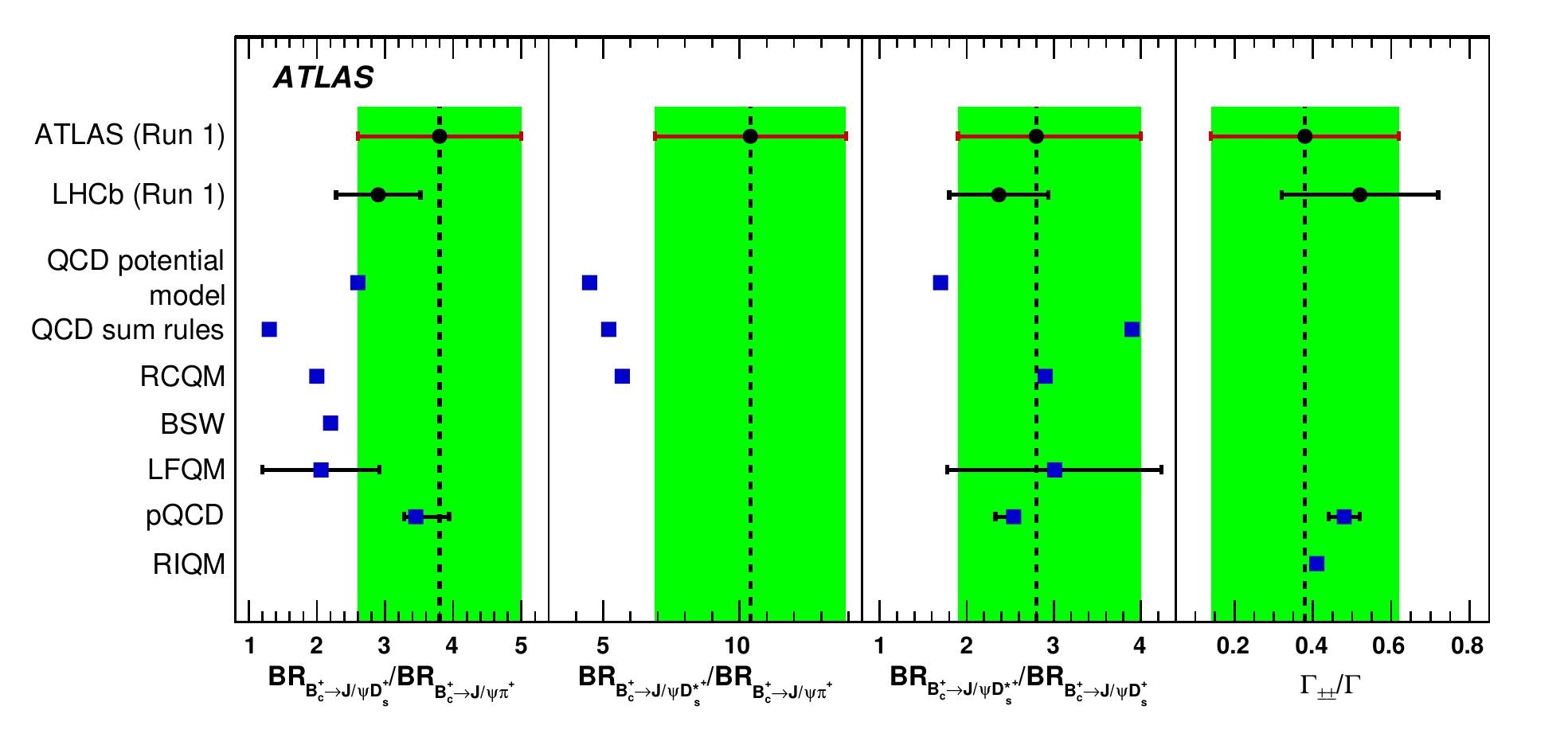}
\caption{Comparison of the results of this measurement with those of 
LHCb~\cite{Aaij:2013gia} and theoretical predictions based on 
a QCD relativistic potential model~\cite{Colangelo:1999zn}, QCD sum 
rules~\cite{Kiselev:2002vz}, relativistic constituent quark model 
(RCQM)~\cite{Ivanov:2006ni}, BSW relativistic quark model 
(with fixed average transverse quark momentum 
$\omega=0.40\GeV$)~\cite{Dhir:2008hh}, 
light-front quark model (LFQM)~\cite{Ke:2013yka}, perturbative QCD 
(pQCD)~\cite{Rui:2014tpa}, and relativistic independent quark model 
(RIQM)~\cite{Kar:2013fna}. The uncertainties of the theoretical predictions are 
shown if they are explicitly quoted in 
the corresponding papers. Statistical and systematic uncertainties added in 
quadrature are quoted for the results of ATLAS and LHCb.
\label{fig:aux:ValComparison}}
\end{figure*}

The measured fraction of the $A_{\pm\pm}$ component 
agrees well with the 
prediction of the relativistic 
independent quark model~\cite{Kar:2013fna} and perturbative 
QCD~\cite{Rui:2014tpa}. 

\section{Conclusion}\label{sec:Conclusion}
A study of $\Bc\to\Jpsi\Ds$ and
$\Bc\to\Jpsi\DsStar$ decays has been performed. The ratios of the branching 
fractions 
$\Br_{\Bc\to\Jpsi\Ds}/\Br_{\Bc\to\Jpsi\pi^+}$, 
$\Br_{\Bc\to\Jpsi\DsStar}/\Br_{\Bc\to\Jpsi\pi^+}$, 
$\Br_{\Bc\to\Jpsi\DsStar}/\Br_{\Bc\to\Jpsi\Ds}$ and the 
transverse polarisation fraction of 
$\Bc\to\Jpsi\DsStar$ decay have been measured by the ATLAS experiment at the
LHC using $pp$ collision data corresponding to an integrated luminosity of 
4.9~\ifb\ at 7~\TeV\ centre-of-mass energy and 20.6~\ifb\ at 8~\TeV. The 
polarisation is found to be well described by the available 
theoretical approaches.
The measured ratios of the branching fraction are generally described 
by perturbative QCD, sum rules, and relativistic quark models.
There is an indication of underestimation of the decay rates for the $\Bc\to\Jpsi\DsOrStar$ 
decays by some models, although the discrepancies do not exceed two standard 
deviations when taking into account only the experimental uncertainty.
The measurement results agree 
with those published by the LHCb experiment.

\section*{Acknowledgements}

We thank CERN for the very successful operation of the LHC, as well as the
support staff from our institutions without whom ATLAS could not be
operated efficiently.

We acknowledge the support of ANPCyT, Argentina; YerPhI, Armenia; ARC,
Australia; BMWFW and FWF, Austria; ANAS, Azerbaijan; SSTC, Belarus; CNPq and FAPESP,
Brazil; NSERC, NRC and CFI, Canada; CERN; CONICYT, Chile; CAS, MOST and NSFC,
China; COLCIENCIAS, Colombia; MSMT CR, MPO CR and VSC CR, Czech Republic;
DNRF, DNSRC and Lundbeck Foundation, Denmark; EPLANET, ERC and NSRF, European Union;
IN2P3-CNRS, CEA-DSM/IRFU, France; GNSF, Georgia; BMBF, DFG, HGF, MPG and AvH
Foundation, Germany; GSRT and NSRF, Greece; RGC, Hong Kong SAR, China; ISF, MINERVA, GIF, I-CORE and Benoziyo Center, Israel; INFN, Italy; MEXT and JSPS, Japan; CNRST, Morocco; FOM and NWO, Netherlands; BRF and RCN, Norway; MNiSW and NCN, Poland; GRICES and FCT, Portugal; MNE/IFA, Romania; MES of Russia and NRC KI, Russian Federation; JINR; MSTD,
Serbia; MSSR, Slovakia; ARRS and MIZ\v{S}, Slovenia; DST/NRF, South Africa;
MINECO, Spain; SRC and Wallenberg Foundation, Sweden; SER, SNSF and Cantons of
Bern and Geneva, Switzerland; NSC, Taiwan; TAEK, Turkey; STFC, the Royal
Society and Leverhulme Trust, United Kingdom; DOE and NSF, United States of
America.

The crucial computing support from all WLCG partners is acknowledged
gratefully, in particular from CERN and the ATLAS Tier-1 facilities at
TRIUMF (Canada), NDGF (Denmark, Norway, Sweden), CC-IN2P3 (France),
KIT/GridKA (Germany), INFN-CNAF (Italy), NL-T1 (Netherlands), PIC (Spain),
ASGC (Taiwan), RAL (UK) and BNL (USA) and in the Tier-2 facilities
worldwide.

\FloatBarrier
\printbibliography

\newpage 
\begin{flushleft}
{\Large The ATLAS Collaboration}

\bigskip

G.~Aad$^{\rm 85}$,
B.~Abbott$^{\rm 113}$,
J.~Abdallah$^{\rm 151}$,
O.~Abdinov$^{\rm 11}$,
R.~Aben$^{\rm 107}$,
M.~Abolins$^{\rm 90}$,
O.S.~AbouZeid$^{\rm 158}$,
H.~Abramowicz$^{\rm 153}$,
H.~Abreu$^{\rm 152}$,
R.~Abreu$^{\rm 30}$
% ...
\bigskip
\\
% ...
$^{11}$ Institute of Physics, Azerbaijan Academy of Sciences, Baku, Azerbaijan\\
% ...
$^{30}$ CERN, Geneva, Switzerland\\
% ...
$^{85}$ CPPM, Aix-Marseille Universit{\'e} and CNRS/IN2P3, Marseille, France\\
% ...
$^{90}$ Department of Physics and Astronomy, Michigan State University, East Lansing MI, United States of America\\
% ...
$^{107}$ Nikhef National Institute for Subatomic Physics and University of Amsterdam, Amsterdam, Netherlands\\
% ...
$^{113}$ Homer L. Dodge Department of Physics and Astronomy, University of Oklahoma, Norman OK, United States of America\\
% ...
$^{151}$ Institute of Physics, Academia Sinica, Taipei, Taiwan\\
$^{152}$ Department of Physics, Technion: Israel Institute of Technology, Haifa, Israel\\
$^{153}$ Raymond and Beverly Sackler School of Physics and Astronomy, Tel Aviv University, Tel Aviv, Israel\\
% ...
$^{158}$ Department of Physics, University of Toronto, Toronto ON, Canada\\
% ...
\end{flushleft}


\end{document}